# Comment on "A Dirac-semimetal two-dimensional BeN$_4$: Thickness-dependent electronic and optical properties [Appl. Phys. Lett. 118, 203103 (2021)]"

Bohayra Mortazavi [a,*], Masoud Shahrokhi [b], Fazel Shojaei [c]

[a] Institute of Photonics, Department of Mathematics and Physics, Leibniz Universität Hannover, Appelstraße 11,30167 Hannover, Germany.
[b] Department of Physics, Faculty of Science, University of Kurdistan, Sanandaj, 66177-15175, Iran.
[c] Department of Chemistry, Faculty of Nano and Bioscience and Technology, Persian Gulf University, Bushehr 75169, Iran.

Corresponding author: *bohayra.mortazavi@gmail.com;

Most recently, Bykov *et al.* [1] reported the successful synthesis of layered structure of BeN$_4$. Followed by the experimental report, Bafekry and coworkers [2] published their first-principles results on the stability, electronic, optical and elastic constants of BeN$_4$ monolayer. In the aforementioned work, authors made numerous wrong claims and reported erroneous results. In the first sentence of abstract, Bafekry *et al.* [2] mentioned that: "*Motivated by the recent experimental realization of a two-dimensional (2D) BeN$_4$ monolayer, …*", which is not true. We emphasis that in the work by Bykov *et al.* [1], the bulk layered BeN$_4$ has been experimentally synthesized and not the BeN$_4$ monolayer. Their theoretical results on the exfoliation energy however confirmed that monolayer of tr-BeN$_4$ is indeed possible. In fact, in our recent theoretical work [3], we predicted an exfoliation energy of 0.32 J/m$^2$ for BeN$_4$ monolayer, which is smaller than the experimentally measured cleavage energy of 0.37 J/m$^2$ for graphite [4].

Bafekry *et al.* [2] were not also careful in preparing the introduction section of their work. For example, they claimed that: "Beryllium (Be)-based 2D materials, such as BeN$_2$,[18,19] Be$_3$N$_2$,[20] BeB$_2$,[21] Be$_5$C$_2$,[22] Be$_2$C,[23] Be$_3$C$_2$,[24] BeS,[25] and BeP$_2$,[26] have been either experimentally realized or theoretically predicted.", nonetheless, none of the cited articles include the experimental synthesis of any of the listed nanosheets. Next, authors discuss the dynamical stability by calculating the phonon dispersion relation and the acquired result is shown in Fig. 1b. First of all, computational details for the reported result, such as the supercell size and K-point mesh are missing. Second, while authors wrote that their density functional theory (DFT) calculations were conducted using the OpenMX code, it appears that this package does not include the possibility to acquire the phonon dispersion relations. Moreover, they reported the phonon dispersion relation along the directions of Γ-M-K-Γ, which belongs to the hexagonal lattices, as those of graphene and h-BN monolayers and thus is not suitable for the analysis of dynamical stability of BeN$_4$ monolayer. Authors later claimed the stability of BeN$_4$ was confirmed the cohesive energy. Nevertheless, in the evaluation of cohesive energy the reference energies for Be and N atoms were considered to be those in the vacuum (isolated). With such a fundamental misunderstanding, even gaseous or liquid forms of BeN$_4$ will be considered energetically stable, because of that fact that with even Van der Waals interactions the cohesive energy will always be positive. Note that N$_2$ is known to be the most stable form of N atoms under ambient conditions. Therefore, for the analysis of energetic stability, the reference energies for consisting atoms should be the ones in their most stable phases and not isolated in vacuum. Although BeN$_4$ monolayer is dynamically and thermally stable [3], but with the presented results authors' conclusion about the stability is not justifiable.

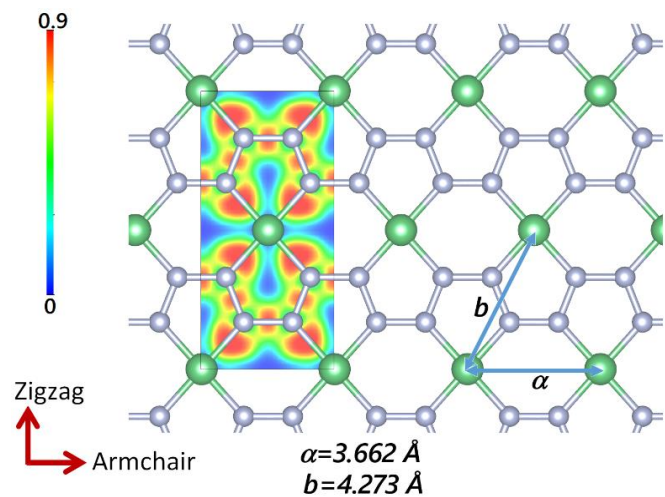

**Fig. 1**, Crystal structure of BeN$_4$ monolayer. In-plane lattice vectors for BeN$_4$ primitive cell are also shown. Contour illustrates electron localization function (ELF) within the rectangular unitcell.

We geometry optimized the rectangular unitcell of BeN$_4$ monolayer and the structure is shown in Fig. 1 (find Appendix for atomic structure). Bafekry *et al.* [2] also erroneously described the crystal lattice of monolayer BeN$_4$ as a hexagonal structure, while the material actually has an oblique primitive cell with in-plane lattice parameters of $|\vec{a}|$ = 3.66 Å, $|\vec{b}|$ = 4.27 Å and γ = 64.64°. Predicted lattice parameters by our DFT matches excellently with those reported in the work by Bykov *et al.* [1]. In Fig. 1, we also plot electron localization function (ELF) [5] to investigate the nature of chemical bonds in BeN$_4$ monolayer. ELF is a topological function and takes a value between 0 to 1. The ELF results shown in Fig. 1 clearly reveal the formation of covalent bonds between Be and N atoms. Bafekry and coworkers [2] also reported the ELF contour in Fig. 1a. However, they did not include the range for the plotted ELF contour. They not only did not include discussions on the plotted ELF contour, but surprisingly stated that "The red (blue) regions indicate high (low) electron density". This confirms that authors do not understand that ELF is a quantitative function and "*electron localization function*" is basically different from "*electron density*".

From the ELF contour shown in Fig. 1, formation of a covalent network in BeN$_4$ monolayer is conspicuous. Moreover, as visible from the atomic structure and also clearly stated in the original work by Bykov *et al.* [1], BeN$_4$ is an anisotropic lattice, meaning that the transport properties, including the Young's modulus and optical properties are dependent on the direction. Authors, surprisingly calculated the Young's modulus and Poisson's ratio of 33.9 GPa and 0.25, respectively, for BeN$_4$ monolayer and concluded that this system is a soft material. As mentioned, reporting a single elastic modulus and Poisson's ratio for an anisotropic material is not accurate, unless the values for different directions are close. More importantly, the calculated elastic properties are highly underestimated, which is very surprising taking into account that DFT-based results are easily reproducible. In the following illustration (Fig. 2), we compare the uniaxial stress-strain relations of BeN$_4$ monolayer along armchair and zigzag directions with that of the graphene. For BeN$_4$ and graphene monolayers we assumed thicknesses of 3.06 Å [1] and 3.35 Å, respectively. As excepted, the BeN$_4$ monolayer shows a highly anisotropic mechanical response. Along the armchair direction, the elastic modulus of BeN$_4$ monolayer is predicted to be ultrahigh and around 945 GPa, which is only by around 5% lower than that predicted for graphene, 996 GPa. In contrast with the work by Bafekry *et al.* [2], we predicted the Poisson's ratio of BeN$_4$ monolayer to be around 0.01. It is thus clear that the reported elastic constants in the work by Bafekry and coworkers are erroneous [2] and basic physics are not observed.

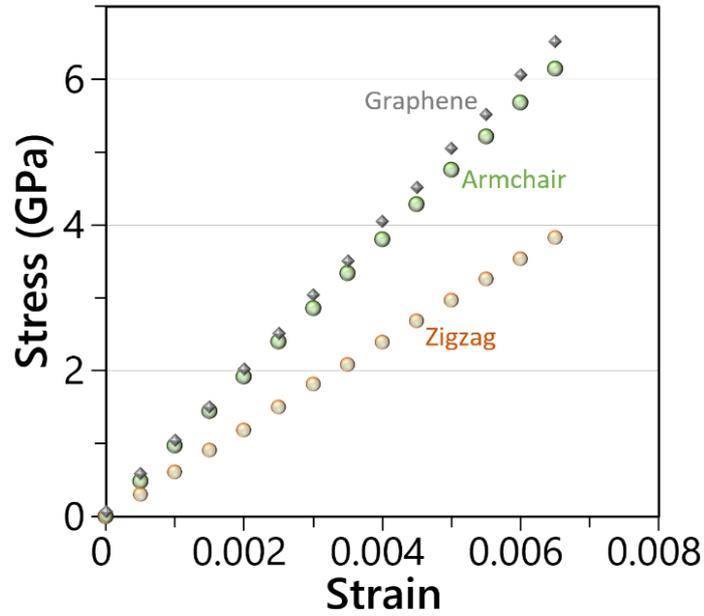

**Fig. 2**, Uniaxial stress-strain responses of BeN$_4$ monolayer along armchair and zigzag directions, compared with graphene.

Later, authors discuss the electronic band structure of single- and multi-layers of BeN$_4$. In comparison with the original results presented by Bykov *et al.* [1], the methodology and presented results for monolayer BeN$_4$ by Bafekry *et al.* [2] does not include any originality and the discussions are clearly superficial. Moreover, details for the first Brillouin zone and corresponding high symmetry points are missing. For the multi-layered BeN$_4$, there exist no information concerning the stacking sequence, and it is not clear if the constructed lattices are at global minimum or not. Worth to note that for metallic systems, both interband and intraband transition contributions are ought to be considered in the evaluation of optical properties [6]. Most importantly, due to the asymmetric lattice of single-layer and bilayer BeN$_4$ along the x- and y-directions (note that x and y directions here correspond to $a$ and *b* lattices in Fig. 1.), the optical properties, likely to mechanical properties can be anisotropic for light polarizations along these axes, which once again has been completely neglected in the work by Bafekry *et al.* [2]. It is also not clear in which direction the optical properties were reported. Moreover, the imaginary part of dielectric function and absorption coefficient of bilayer BeN$_4$ start with a gap of ~0.25 eV, which intrinsically correspond to the semiconducting property. The reported dielectric constant value (the real part of dielectric function at E = 0) Bafekry *et al.* [2] is ~4 for 2L BeN$_4$ monolayer, which is very small for a semi-metal compound. Such as low dielectric constant is generally observable for a wide band gap semiconductor.

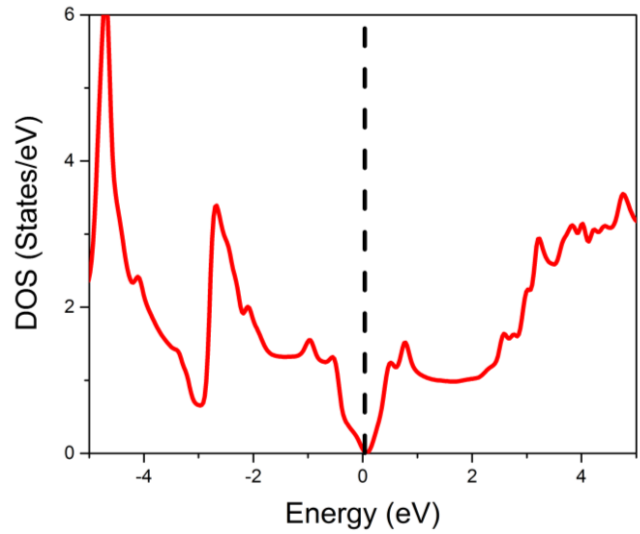

**Fig. 3**, DOS of bilayer BeN$_4$. The zero energy is set to the Fermi level.

We calculated the electronic DOS and optical properties for bilayer BeN$_4$ with bulk stacking pattern as shown in Fig. 3. The real part of dielectric function (Re ε), the imaginary part of dielectric function (Im ε) and the absorption coefficient for 2L BeN$_4$ are shown in Fig.4. It is obvious the optical properties for bilayer BeN$_4$ are anisotropic along x- and y-axes. The dielectric constant values along x- and y-axes are 10.5 and 19.5 which are much greater than that reported by Bafekry *et al.* [2]. Interestingly, both imaginary part of dielectric constant and absorption coefficients starts without any gap.

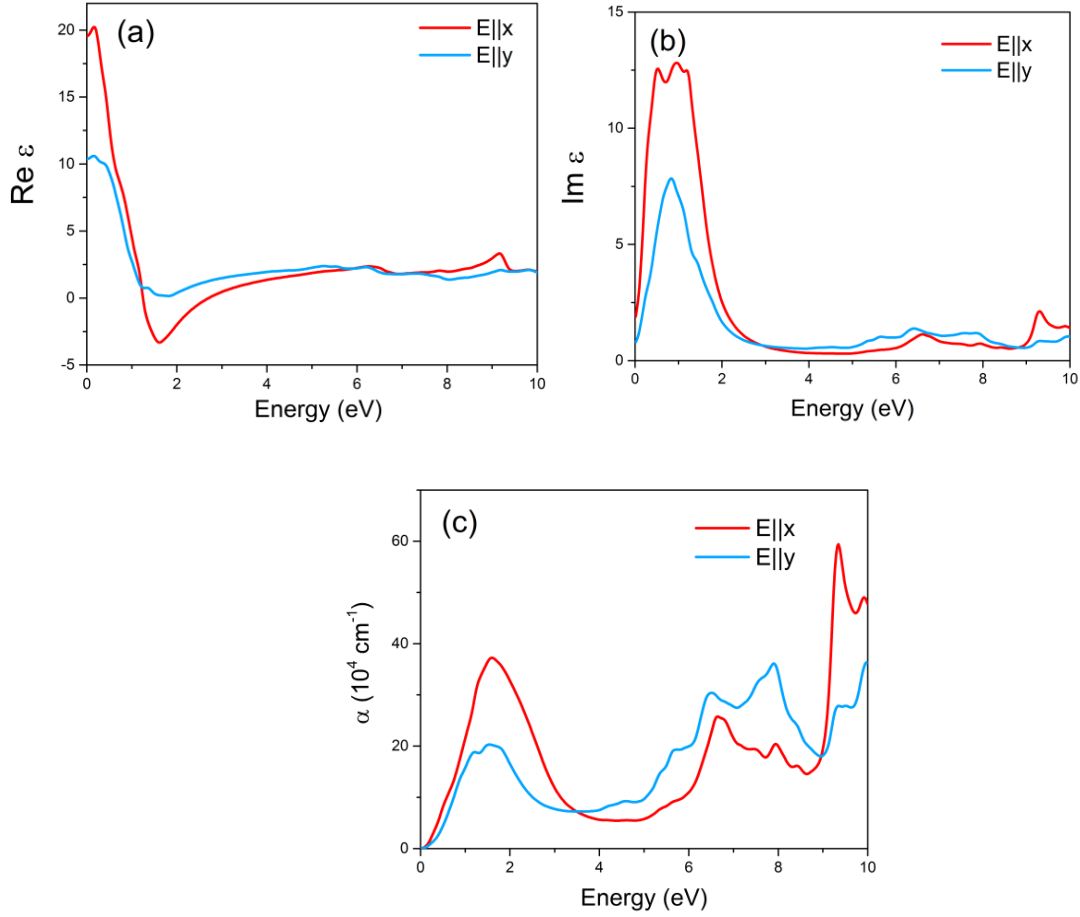

**Fig. 4**, (a) Real part of the dielectric constant, (b) the imaginary part of the dielectric constant and (c) the absorption coefficient of bilayer BeN$_4$ along x - and y-axes. Please note that x and y directions here correspond to $a$ and $b$ lattices in Fig. 1.

Based on the above discussions, the work by Bafekry *et al.* [2] is an erroneous study and includes major errors in referring to the original experimental work by Bykov *et al.* [1] and studies in the literature, providing the computational details, analyzing the dynamical and energetic stability, calculating the elastic constants, evaluating the optical properties and understanding the basic physics of electron localization function.

*Methods*

DFT calculations were performed with the generalized gradient approximation (GGA) and Perdew–Burke–Ernzerhof (PBE) [7], as implemented in *Vienna Ab-initio Simulation Package* [8,9]. Projector augmented wave method was used to treat the electron-ion interactions [10,11]. We applied periodic boundary conditions in all directions with a 20 Å vacuum layer to avoid image-image interactions along the monolayer's thickness. We considered a cutoff energy of 600 eV for the plane waves. For the geometry optimization of the rectangular unitcell, atoms and lattice were relaxed according to the Hellman-Feynman forces using conjugate gradient algorithm until atomic forces drop to lower than 0.001 eV/Å [12]. The first Brillouin zone (BZ) was sampled with 12×8×1 Monkhorst-Pack [13] k-point grid. Mechanical properties were examined by conducting uniaxial tensile simulations over. VESTA package [14] was used to plot the atomic structure and ELF contour as well. DFT-D3 [15] van der Waals dispersion correction is considered for the modeling of bilayer structure. For optical properties we used primitive unitcell and for calculations we used 22×18×1 Monkhorst-Pack [13] k-point grid. The imaginary part of the interband dielectric permittivity is given by [6]:

$$\mathrm{Im}\,\varepsilon_{\alpha\beta}(\omega) = \frac{4\pi^2 e^2}{\Omega} \lim_{q \to 0} \frac{1}{|q|^2} \sum_{c,v,k} 2w_k \delta(\varepsilon_{ck} - \varepsilon_{vk} - \omega) \times \langle u_{ck+e_\alpha q} | u_{vk} \rangle \langle u_{ck+e_\beta q} | u_{vk} \rangle^* \quad (1)$$

where $q$ is the Bloch vector of the incident wave, $w_k$ is the k-point weight and the band indices $c$ and $v$ are restricted to the conduction and the valence band states, respectively. By using the $\mathrm{Im}\,\varepsilon_{\alpha\beta}(\omega)$, one can determine the corresponding real part via the Kramers–Kronig relations:

$$\mathrm{Re}\,\varepsilon_{\alpha\beta}(\omega) = 1 + \frac{2}{\pi} P \int_0^\infty \frac{\omega' \mathrm{Im}\,\varepsilon_{\alpha\beta}(\omega')}{(\omega')^2 - \omega^2 + i\eta} d\omega' \quad (2)$$

where $P$ denotes the principle value and $\eta$ is the complex shift.

The adsorption coefficient determined as:

$$a_{\alpha\beta}(\omega) = \frac{2\omega k_{\alpha\beta}(\omega)}{c} \quad (3)$$

where $k_{\alpha\beta}$ is imaginary part of the complex refractive index and $c$ is the speed of light in vacuum, known as the extinction index. It is given by the following relations

$$k_{\alpha\beta}(\omega) = \sqrt{\frac{|\varepsilon_{\alpha\beta}(\omega) - \mathrm{Re}\,\varepsilon_{\alpha\beta}(\omega)|}{2}} \quad (4)$$

The reflectivity is given by

$$R_{ij}(\omega) = \frac{(n-1)^2 + k^2}{(n+1)^2 + k^2} \quad (5)$$

where $n$ and $k$ are real and imaginary parts of the complex refractive index, which are known as the refractive index and the extinction index, respectively.

## Acknowledgment

B.M. and X.Z. appreciate the funding by the Deutsche Forschungsgemeinschaft (DFG, German Research Foundation) under Germany's Excellence Strategy within the Cluster of Excellence PhoenixD (EXC 2122, Project ID 390833453). F.S. thanks the Persian Gulf University Research Council, Iran for support of this study. B. M is greatly thankful to the VEGAS cluster at Bauhaus University of Weimar for providing the computational resources.

## Appendix

Atomic positions for the rectangular unitcell of BeN$_4$ monolayer in VASP POSCAR format

```
BeN4-Rec
  1.00000000000000
    3.6618114570060962   0.0000000000000000   0.0000000000000000
    0.0000000000000000   7.7224544425148238   0.0000000000000000
    0.0000000000000000   0.0000000000000000  20.0000000000000000
   N    Be
   8    2
Direct
 0.3213218618313992  0.1702513769760617  0.5000000000000000
 0.8212477474301051  0.6702542072785187  0.5000000000000000
 0.6881212205488810  0.1702536823538319  0.5000000000000000
 0.1880470397073434  0.6702507888373577  0.5000000000000000
 0.8213219299981124  0.3316676051419731  0.5000000000000000
 0.3212478006664892  0.8316646331463877  0.5000000000000000
 0.1881222242215886  0.3316641023320760  0.5000000000000000
 0.6880481399660080  0.8316670599750109  0.5000000000000000
 0.0046825160753059  0.0009590406566673  0.5000000000000000
 0.5046803135547648  0.5009589693021326  0.5000000000000000
```

## References

[1]  M. Bykov, T. Fedotenko, S. Chariton, D. Laniel, K. Glazyrin, M. Hanfland, J.S. Smith, V.B. Prakapenka, M.F. Mahmood, A.F. Goncharov, A. V. Ponomareva, F. Tasnádi, A.I. Abrikosov, T. Bin Masood, I. Hotz, A.N. Rudenko, M.I. Katsnelson, N. Dubrovinskaia, L. Dubrovinsky, I.A. Abrikosov, High-pressure synthesis of Dirac materials: Layered van der Waals bonded BeN4 polymorph, Phys. Rev. Lett. 126 (2021) 175501. https://doi.org/10.1103/PhysRevLett.126.175501.

[2]  A. Bafekry, C. Stampfl, M. Faraji, M. Yagmurcukardes, M.M. Fadlallah, H.R. Jappor, M. Ghergherehchi, S.A.H. Feghhi, A Dirac-semimetal two-dimensional BeN4: Thickness-dependent electronic and optical properties, Appl. Phys. Lett. 118 (2021) 203103. https://doi.org/10.1063/5.0051878.

[3]  B. Mortazavi, F. Shojaei, X. Zhuang, Ultrahigh stiffness and anisotropic Dirac cones in BeN4 and MgN4 monolayers: A first-principles study, Mater. Today Nano. (2021) 100125. https://doi.org/10.1016/j.mtnano.2021.100125.

[4]  W. Wang, S. Dai, X. Li, J. Yang, D.J. Srolovitz, Q. Zheng, Measurement of the cleavage energy of graphite, Nat. Commun. (2015). https://doi.org/10.1038/ncomms8853.

[5]  B. Silvi, A. Savin, Classification of Chemical-Bonds Based on Topological Analysis of Electron Localization Functions, Nature. 371 (1994) 683–686. https://doi.org/10.1038/371683a0.

[6]  F. Wooten, Optical properties of solids, Academic press, 2013.

[7] J. Perdew, K. Burke, M. Ernzerhof, Generalized Gradient Approximation Made Simple., Phys. Rev. Lett. 77 (1996) 3865–3868. https://doi.org/10.1103/PhysRevLett.77.3865.

[8] G. Kresse, J. Furthmüller, Efficient iterative schemes for ab initio total-energy calculations using a plane-wave basis set, Phys. Rev. B. 54 (1996) 11169–11186. https://doi.org/10.1103/PhysRevB.54.11169.

[9] J.P. Perdew, K. Burke, M. Ernzerhof, Generalized Gradient Approximation Made Simple, Phys. Rev. Lett. 77 (1996) 3865–3868. https://doi.org/10.1103/PhysRevLett.77.3865.

[10] P.E. Blöchl, Projector augmented-wave method, Phys. Rev. B. 50 (1994) 17953–17979. https://doi.org/10.1103/PhysRevB.50.17953.

[11] G. Kresse, D. Joubert, From ultrasoft pseudopotentials to the projector augmented-wave method, Phys. Rev. B. 59 (1999) 1758–1775. https://doi.org/10.1103/PhysRevB.59.1758.

[12] G. Kresse, J. Hafner, Ab initio molecular dynamics for liquid metals, Phys. Rev. B. 47 (1993) 558–561. https://doi.org/10.1103/PhysRevB.47.558.

[13] H. Monkhorst, J. Pack, Special points for Brillouin zone integrations, Phys. Rev. B. 13 (1976) 5188–5192. https://doi.org/10.1103/PhysRevB.13.5188.

[14] K. Momma, F. Izumi, VESTA 3 for three-dimensional visualization of crystal, volumetric and morphology data, J. Appl. Crystallogr. 44 (2011) 1272–1276. https://doi.org/10.1107/S0021889811038970.

[15] S. Grimme, J. Antony, S. Ehrlich, H. Krieg, A consistent and accurate ab initio parametrization of density functional dispersion correction (DFT-D) for the 94 elements H-Pu, J. Chem. Phys. 132 (2010) 154104. https://doi.org/10.1063/1.3382344.